\documentstyle[11pt,aaspp4]{article} 
\newcommand{\kms}{km~s$^{-1}$}

\newcommand{\etal}{{\it et al.\/}}

\begin{document}

\title{The Kinematics of the Outer Halo of M87\altaffilmark{1}}

\author{Judith G. Cohen\altaffilmark{2}}

\altaffiltext{1}{Based in large part on observations obtained at the
	W.M. Keck Observatory, which is operated jointly by the California 
	Institute of Technology and the University of California}
\altaffiltext{2}{Palomar Observatory, Mail Stop 105-24,
	California Institute of Technology, Pasadena, CA \, 91125
        jlc@astro.caltech.edu}

\begin{abstract}

Radial velocities are presented for a new sample of globular clusters
in the outer halo of M87 at a distance of 300 to 540 arcsec (24 to 43 kpc)
from the center of this galaxy.  These are used to augment our previously
published data and an analysis of the rotation and velocity dispersion
of the M87 globular cluster system is carried out.  The rotation
is $\sim300$ \kms\ at $R = 32$ kpc, at which point the velocity
dispersion is also still quite high, $\sim450$ \kms.   The high
rotation is interesting. The outer halo
of M87 is, as was found in our previous kinematic analysis, very massive. 

\end{abstract}


\section{Introduction}

Cohen \& Ryzhov (1997) (henceforth CR) present an analysis of the kinematics
of M87 based on radial velocities for 205 globular clusters 
(henceforth GCs) in the
halo of M87.  In that paper we adopted a rotation curve which rises
linearly with radius from the center of M87 ($R$) until reaching 180 \kms\ at a radius
of 225 arcsec.
Beyond that point, and particularly beyond $R \sim 350$ arcsec,
our sample became too sparse to establish
a meaningful rotation curve, and, following the example of
the Galaxy, we adopted the fixed value of
180 \kms\ as the rotation of the outer part of M87.

The rotation of M87 is important for two reasons.  The total
angular momentum of M87 is
a clue to the mode of formation of this massive elliptical galaxy
at the center of the Virgo cluster cooling flow. This parameter can
help distinguish between formation via a gravitational collapse
as advocated by Peebles (1969) as compared to a merger model.
In addition, a major goal of CR
was the determination of the enclosed
mass of M87 as a function of $R$.  
A proper dynamical mass estimate requires knowledge of both
rotation and dispersion.

Kissler-Patig \& Gebhardt (1998) suggested, based on a reanalysis
of our previously published data, that the rotation in the outer
part of M87 is large and decreases the
observed $\sigma_v$ more than we allowed for in our previous work.  
Since our sample of GCs there is very sparse,
the purpose of this paper is to present radial velocities
for an additional group of M87 globular clusters selected
to be in spatial positions which
maximize their ability to constrain the rotation of the outer halo
of M87.  We then
combine this new data with our previously published data to re-examine
the issue of the rotation and velocity dispersion in the outer part of M87.

Whitmore \etal\ (1995) derive a distance to M87 of 17 Mpc from
the turnover of the M 87 GC luminosity function.  This is in
excellent agreement with the mean of the Cepheid distances for Virgo cluster
spirals from
Pierce \etal\ (1994) and from Ferrarese \etal\ (2000) 
of 16 Mpc, ignoring NGC 4639 (Saha \etal\ 1996) as 
a background object.  
Hence we adopt 16.3 Mpc (80 pc/arcsec) as the distance to M87. 

\section{New Observations of M87 GCs}

We observed a single slitmask of M87 GC candidates with the
Low Resolution Imaging Spectrograph (henceforth LRIS) (Oke \etal\ 1995)
at the Keck Telescope during the spring of 1999.  This mask
was designed to include objects located along the major
axis of M87 $\approx$400 arcsec SE of the nucleus, 
with the slit length running perpendicular
to the major axis.  It contains objects with $R \sim300$
to 540 arcsec from the
center of M87.  Candidates from the catalog of Strom \etal\ (1981) 
in this area were checked on existing direct images to determine
if they appeared to be M87 GCs.  Those 
that are brighter than
$B = 22$ mag and that were not observed by CR were included here.
In addition, our new sample is at the maximum radius from M87
of the Strom \etal\ survey.  Since we wish to
include M87 GCs at even larger radii,
we selected several stellar objects from the area on these
LRIS images beyond that of
the Strom \etal\ survey which appeared to be
M87 GCs. Finally, we included
the object in this region with the most discordant $v_r$ 
from CR.

The slitmask was observed with the same instrumental configuration
of LRIS as was used in CR, but the grating was tilted to center
the spectra at H$\alpha$.  Only H$\alpha$ was used to determine
the radial velocities.  Because of the plethora of night sky
lines and the use of a 1 arcsec wide slit, these $v_r$ are more
accurate than those of CR, with typical 1$\sigma$ errors of only $\pm$50 \kms.
Table 1 lists the new radial velocities.  There are 20
entries, four of which are included in CR.  Two of the M87 GCs
were found through the procedure described above
and have not been previously cataloged.  
They are assigned identifying
numbers beginning with 6000;  their location
and brightness in the $R$ filter bandpass using the standard
stars of Landolt (1992) are given in footnotes to the table.
Both are more than 500 arcsec from the center of M87.

The $v_r$ from CR for the four cases in common are given in the last column
of Table~1.  The agreement is very gratifying and
suggests yet again that the quoted error estimates for these GC
radial velocities are realistic.

Several interlopers were also found among these new spectroscopic observations.
Strom 81 is a galaxy showing strong H$\alpha$
emission with $z = 0.335$, and Strom 87, 221 and 286
are galactic stars.

\section{Rotation and Velocity Dispersion Analysis}

The sample of 16 new GCs in M87 
presented in Table~1 more than
doubles the sample with spectroscopic $v_r$ with $R > 420$ arcsec.
This should produce credible measures of the rotation and
velocity dispersion in the outer part of M87.
The sample of M87 GC radial velocities we utilize below
is that of CR as augmented and
updated in Cohen, Blakeslee \& Ryzhov (1998) plus the new material
presented in Table~1.  This gives a total of 222 objects believed
to be M87 GCs.

\subsection{Qualitative Results}

To demonstrate in a simple yet convincing manner that significant
rotation exists in the outer halo of M87,
we assume that the rotation is about a fixed position angle,
that characteristic of the isophotes of M87.  A modern study
of the isophotes of M87 by Zeilinger, Moller \& Stiavelli (1993)
finds the major axis of the galaxy to be at PA = 160$^{\circ}$
and the effective radius to be $\sim90$ arcsec (7.2 kpc).  The
ellipticity they deduce increases with radius outside the core, reaching 0.2
at 1.3$r_{eff}$.  (In early work on this galaxy, Cohen 1986
found  $\epsilon \sim 0.2$ at $R = 230$ arcsec with
a position angle of 155$^{\circ}$.)
McLaughlin, Harris \& Hanes (1994) establish that the PA of the M87 GC
system is identical to that of the underlying galaxy halo light.
Kundu \etal\ (1999) find the PA of
the globular cluster system very close to the center of M87
to be somewhat larger, $\sim190^{\circ}$.
The analysis of the isophotes of this galaxy by
Blakeslee (1999) covers this entire range of $R$ and shows the twisting
of their major axis 
within the central 20 arcsec and the increase of $\epsilon$ outward.

We further assume that the mean velocity at all radii is the
systemic velocity of M87, 1277$\pm$5 \kms\ (van der Marel 1994).  
Figure 1 shows the rotation curve deduced under these assumptions for
all M87 GCs beyond $R = 380$ arcsec with separate symbols
for the new observations presented here and for the published
data of CR.  The horizontal line indicates the systemic velocity 
while the curve shown is
for $v_{rot}sin(i) = 300$ \kms.
The sample at such large distance from the center of M87 is now
reasonably large (31 M87 GCs) with good coverage near
PA = 160 and 340$^{\circ}$, i.e. along the major axis both towards
the SE and the NW from the center of M87, and clearly demonstrates that
the outer part of M87 is rotating rapidly.  In addition, Figure~1 shows that
the velocity dispersion of the M87 GC system
is still very large, $\sim 400$ \kms, even at the
outermost point reached. 

\subsection{Quantitative Determination of the Rotation Curve of M87}

Both the analysis given in CR and that of Kissler-Patig \& Gebhardt (1998)
suggest that the assumptions made above with respect to the position
angle of the axis of rotation and the systemic velocity are correct,
and we therefore adopt them for our detailed analysis of the rotation
of M87.  

GCs that are located on the minor
axis constrain the mean velocity and velocity dispersion but
contribute no information towards
determining the amplitude of rotation.
Having made the above assumptions, given the large velocity dispersion of the
M87 GC system compared to the expected rotation, the GCs near the minor axis
of M87 contribute mostly noise to the determination of the amplitude
of rotation.  We thus do not include clusters with $|cos(\theta-\theta_0)| < 0.30$
in the rotation solution, the choice of angle being somewhat arbitrary
but based on the ratio of $\sigma_v/v_{rot}$.
This excludes GCs in two arcs, each $35\deg$ long, centered on each end of
the minor axis, or 19\% of the total sample, if the GCs are distributed
uniformly in angle at all radii.  All GCs, including the ones
rejected here, are subsequently utilized to determine the velocity dispersion.

The rotation analysis is thus reduced to finding a suitable statistically
accurate representation of the set of values
$ \{v_{rot}(R)\}_i = (v_r(i) - v_{sys})/cos(\theta-\theta_0)$ 
for GCs within a specified range of $R$.  Within each radial bin considered,
a two step $\chi^2$ minimization solution was implemented to solve
for an appropriate value of the amplitude of the rotation, $v_{rot}(R)$.
The errors in the 
individual terms on the right hand side of the above expression
for a constant observational uncertainty in $v_r$ are highly variable
and depend on  $\theta$.  The procedure adopted
allows for these varying uncertainties.
In the first pass, an initial guess at $\sigma_v(R)$ is used, and a 
solution for $v_{rot}(R)$
is found.   This solution for $v_{rot}(R)$ is used to derive $\sigma_v(R)$.
Since $\sigma_v(R)$ is used is used as the error estimate
for each point for $v_{rot}(R)$, a second pass solution, which only makes
very small updates,
is then carried out to derive $v_{rot}(R)$.

The GCs in the spectroscopic sample are sorted in ascending order in $R$.
The analysis is carried out with 30 point bins at the extreme inward
and outward points, increasing to bins with 50 GCs wherever possible.
The bin center is shifted outward by 1 GC, and the solution is repeated.
Figure~2 displays the resulting solution for $v_{rot}(R_m)$ as a 
function of the
median distance $R_m$ from the center of M87 for the GCs in each bin.  
The errors are calculated assuming Gaussian statistics
within each bin.  The radial extents for a few typical bins are 
indicated by the horizontal lines in the Figure.  The rotation curve
is heavily oversampled; there are 
only five independent points on this Figure.

\subsection{Determination of the Velocity Dispersion}

The calculation of the velocity dispersion requires removal of
the rotational velocity.  This is done using
a smoothed version of
the rotation amplitude
found as described above.  The 
biweight estimator described in Beers, Flynn
\& Gebhardt (1990) which is strongly resistant to outliers is used. 
The instrumental contribution to $\sigma_v(R)$ is also removed in
quadrature.
The same variable binning with $R$ used for the rotation solution is
adopted here.  The entire spectroscopic sample 
of M87 GCs is used.

The resulting radial profile of $\sigma_v$ is shown in
Figure~2.  The radial extent 
of a few typical bins is 
indicated by the horizontal lines in the Figure.
As is the case for Figure~1, the velocity dispersion profile 
shown in this Figure is heavily oversampled.

\section{Discussion}

\subsection{Comments on Galaxy Formation}

Both the qualitative and the quantitative analysis show that the
outer part of the halo of M87 with $R \sim 400$ arcsec, 32 kpc
from the center of M87 and at 4.4$r_{eff}$, is rotating 
with a projected rotational
velocity of $\sim300$ \kms.  This is a very large rotational
velocity to be found so far out in M87, implying that
the total angular momentum of M87 is very large.  It is
interesting to note that the halo population of the Galactic
globular cluster system shows a rotation of only 50 $\pm$23 \kms\ 
while the disk population of Galactic GCs is highly flattened
with $v_{rot}$ 152 $\pm29$ \kms\ (Zinn 1985).

Among the various theories of galaxy formation, there are
several that are often applied to elliptical galaxies.
The theory of dissipationless collapse from a single gas cloud
through a gravitational instability, and acquisition of angular
momentum through tidal torques 
was worked out by Peebles (1969).  Binney (1978) calculated
the expected rotation velocities of elliptical galaxies
under various assumptions regarding orbit anisotropy.
Detailed N-body simulations
along these lines have been carried out by several groups,
including Barnes \& Efstathiou (1987),
Stiavelli \& Sparke (1991) and Ueda \etal\ (1994).
The relevant parameter for comparison of observations
with analytical and numerical
models of galaxy formation is the spin parameter (Peebles 1971),
a dimensionless combination of the
total angular momentum, total mass, and total energy for a galaxy,
$\lambda = J E^{1/2} G^{-1} M^{-5/2}$.  This does not
exceed 0.1 for such models, while Kissler-Patig \& Gebhardt (1998)
find $\lambda \sim 0.18$ for M87, with the caution that calculating
$\lambda$ from our existing observational material requires
an extrapolation to the half mass radius in the halo of M87, much
further out in $R$.  
The complication of possible triaxial
shapes rather than oblate isotropic rotators, reviewed
by de Zeeuw \& Franx (1991), further obscures the validity of
such calculations.

The other major theory currently in vogue 
for the formation of elliptical galaxies is through the
merger of several large gas rich fragments, as
originally suggested
by Toomre \& Toomre (1972).  Such models come in various flavors,
with the mergers happening relatively late and the pieces being 
entire galaxies as in Kormendy (1989) and Kormendy \& Sanders (1992)
or happening at early times involving protogalaxy clumps
as in White \& Rees (1977).  Recent numerical simulations
of this type of model, expanding on the earlier work of Barnes (1988),
can be found in Hernquist \& Bolte (1993) and in Bekki (1998).
It is likely that such a model is more capable of reproducing 
the high rotation we find.  Furthermore the observation of non-symmetric
diffuse light in the halo of M87 extending 100 Kpc from its
center by Weil, Bland-Hawthorn \& Malin (1997) may also be explained
by the recent accretion of a low mass galaxy.

\subsection{The Effect of the Present Results on Those of Cohen
\& Ryzhov (1997)}

Even with the large rotation we have found in the outer part of M87,
we find the velocity dispersion after the observed values
are corrected for the rotation to be still high there. 
$\sigma_v$ in the outermost bin
of CR agrees to within 5\% with the value found here.  Only
in one of the the nine
radial bins used by CR is $\sigma_v$ from the current solution
smaller than
the values given in CR.  We therefore expect that the
results of CR with respect to the distribution
of mass within M87 still are approximately correct.

\section{Summary}

We have presented spectroscopic observations of a new sample of M87
GCs chosen to put maximum leverage on a determination of the rotation
in the outer halo of M87.  Using this data combined with our
previously published data we find that the rotation of M87
increases outward and reaches a value of $\sim$300 \kms\ at
a distance of 400 arcsec (32 kpc, 4.4 $r_{eff}$) from the center
of this galaxy, confirming the suggestion based on our previously
published set of M87 GC radial velocities by Kissler-Patig
\& Gebhardt (1998) of high rotation.
That is rather surprising.  The velocity dispersion
remains high even at that large $R$, and the enclosed 
total mass is very large, as our
earlier work (see CR) suggested.  

The high rotation may provide
a clue to the mode of formation of M87, but  
the calculation of total spin is uncertain as it
involves extrapolation to still larger radii.  In addition, M87 is
a located in a very special place, the center of a very large
mass concentration, a large cooling flow, etc.  Its history
may be quite different from that of most ellipticals, even of
most massive ellipticals.

The results of the initial spectroscopic studies of the dynamics
of elliptical galaxies
in the early 1980s were very surprising.  Davies \etal\ (1983)
showed that low luminosity ellipticals
rotate as rapidly as spiral bulges and as rapidly as predicted by
models with oblate figures and isotropic distributions of residual
velocities.  However, as was discovered by Illingworth (1977), 
high luminosity ellipticals show surprisingly
small values of $v_{rot}/\sigma_v$.
Our results for the outer part of M87, where $v_{rot}/\sigma_v \sim 0.6$,
include no
correction to $v_{rot}$ for projection effects, which would only
make this ratio larger.  Similar suggestions for
high rotation in the outer parts of luminous ellipticals near the
center of large clusters of galaxies have been obtained from 
the analysis of a sample
of globular clusters in M49 by Sharples \etal\ (1998) and one in
NGC 1399 by Kissler-Patig \etal\ (1999).  This
gives rise to some interesting questions.  Is this
large rotation in the outer parts of the M87 GC system also shared
by the M87 stellar halo ?  How far out does this rotation continue ?
To be provocative, we might ask if most luminous elliptical galaxies have $v_{rot}/\sigma_v > 0.5$ in
their outer parts, and whether this was missed in earlier studies due
to limitations on slit length and on surface brightness in 
those spectroscopic studies ?

One of the few ways to explore these issues is to attempt to find
large samples of GCs even further out in the halo of M87.  We have
paved the way with a first identification of M87 GCs beyond the
spatial limit of existing surveys.

\acknowledgements The entire Keck/LRIS user community owes a huge debt
to Jerry Nelson, Gerry Smith, Bev Oke, and many other people who have
worked to make the Keck Telescope and LRIS a reality.  We are grateful
to the W. M. Keck Foundation, and particularly its late president,
Howard Keck, for the vision to fund the construction of the W. M. Keck
Observatory.  We thank John Blakeslee and Patrick C\^ot\'e for helpful
discussions.

\clearpage

%
%
\begin{deluxetable}{lrrr}
\tablenum{1}
\tablewidth{0pt}
\tablecaption{New Radial Velocity Measurements for M87 Globular Clusters}
\label{tab1}
\tablehead{\colhead{ID\tablenotemark{a}} & \colhead{$R$\tablenotemark{b}} & 
\colhead{$v_r$} & \colhead{$v_r$ from CR} \nl
\colhead{} &  \colhead{(arcsec)} & \colhead{(\kms)} & \colhead{(\kms)} \nl
\colhead{} & \colhead{} & \colhead{($\pm50$ \kms)} &
\colhead{($\pm100$ \kms)} \nl
}
\startdata
16 &  530 & 1115 \nl
28 &  494 & 1157 \nl
41 &  486 & 1814 \nl
66  & 517 & 2212 & 2260 \nl
77  & 367 & 1774 \nl
94  & 393 & 963 \nl
103  & 375 & 1092 \nl
137  & 367 & 1764 \nl
176  & 502 & 2167 & 2210 \nl
177  & 479 & 1540 & 1629 \nl
300  & 455 & 1977 \nl
314  & 309 & 1167 & 1236 \nl
379  & 413 & 2244 \nl
415  & 395 & 1817 \nl
477  & 363 & 1584 \nl
514  & 412 & 1617 \nl
603  & 397 & 1741 \nl
731  & 383 & 960 \nl
6003\tablenotemark{c}  & 516 & 1741 \nl
6004\tablenotemark{d}  & 540 & 1741 \nl
\enddata
\tablenotetext{a}{ID from Strom \etal\ (1981) except as noted.}
\tablenotetext{b}{Distance from the center of M87.}
\tablenotetext{c}
{New M87 GC located 2.7 arcsec W and 27.3 arcsec S of Strom 176, $R$ = 21.05 mag}
\tablenotetext{d}{New M87 GC
located 1.2 arcsec W and 54.4 arcsec S of Strom 176, $R$ = 21.22 mag }
\end{deluxetable}

\clearpage

\begin{figure}
\epsscale{0.7}
\plotone{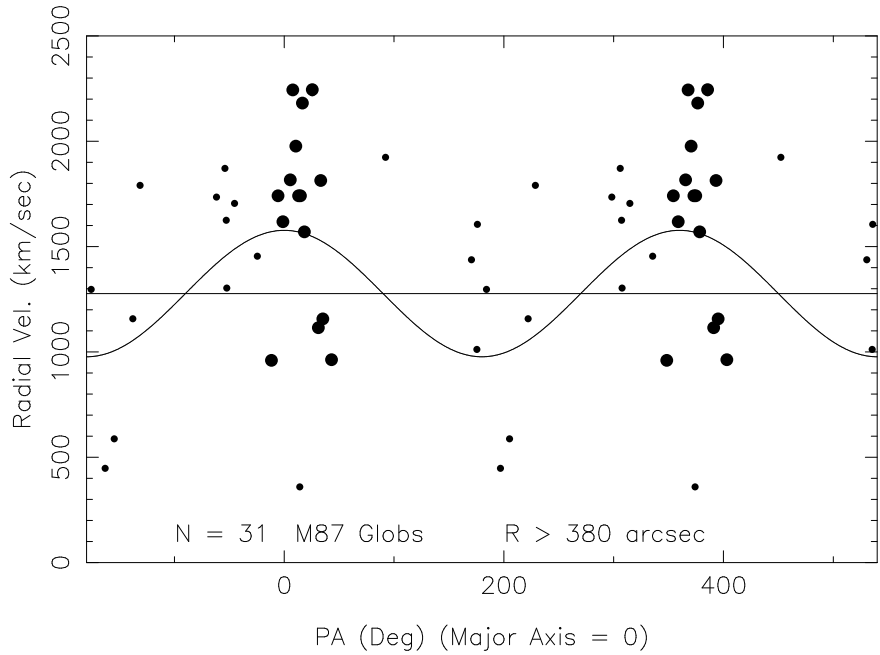}
\caption[figure1.ps]{The radial velocity of M87 GCs with $R > 380$ arcsec
is shown as a function of position angle about the major axis of
M87.  The large filled circles are from the data of Table~1, while
the smaller filled circles represent M87 GCs from CR. 
The curve shows the projected rotation as a function of
$\theta$ with amplitude of 
300 \kms\ about the systemic
velocity (indicated by the horizontal line).
\label{fig1}}
\end{figure}

\begin{figure}
\epsscale{0.6}
\plotone{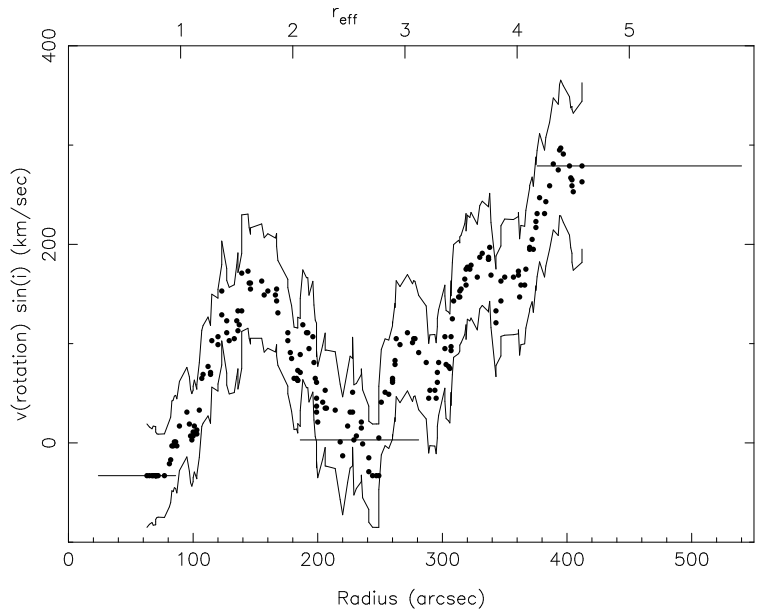}
\caption[figure2.ps]{The projected rotational velocity inferred for
M87 from its system of GCs is shown as a function of $R$. Each point
represents a group of GCs at similar radii.  The range of
radii included within three typical
bins are indicated by the horizontal lines.  The $1\sigma$ 
error bars for each point are also indicated by the lines.
Details of the binning scheme are described in the text.
\label{fig2}}
\end{figure}

\begin{figure}
\epsscale{0.7}
\plotone{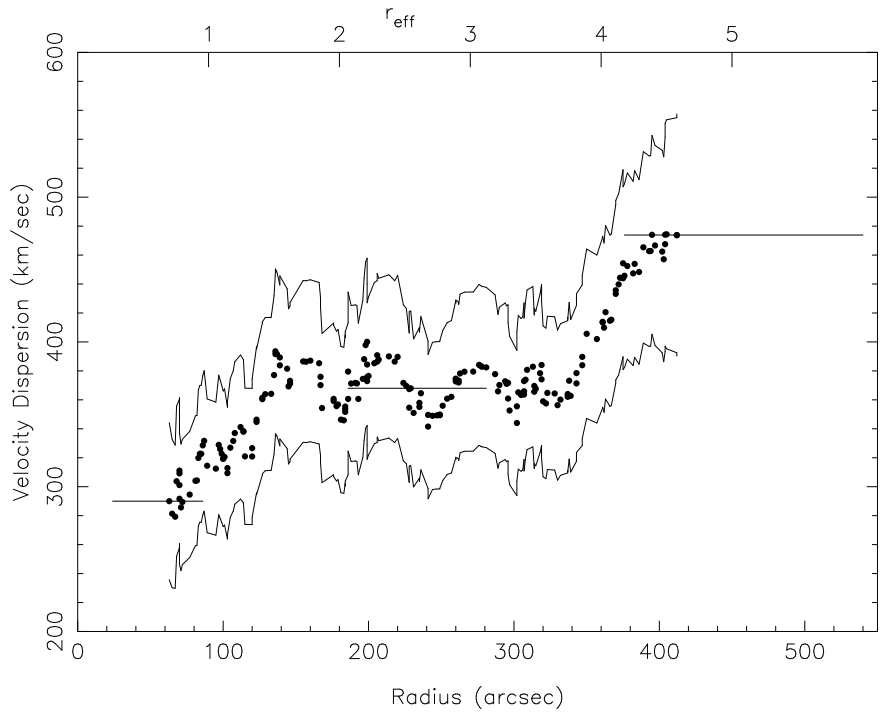}
\caption[figure3.ps]{The same as Figure~2 for the velocity dispersion.
The contribution of rotation to the observed $\sigma_v$ has been removed.
\label{fig3}}
\end{figure}

\end{document}